\documentclass[a4paper,aps,prd,10pt,preprintnumbers,showpacs,twocolumn,superscriptaddress,nofootinbib,amsmath,amssymb,floatfix]{revtex4-1}
\usepackage{graphicx}
\usepackage{cmap}
\usepackage[utf8]{inputenc}
\usepackage[T1]{fontenc}
\makeatletter
\def\@bibdataout@aps{%
 \immediate\write\@bibdataout{%
  @CONTROL{%
   apsrev41Control%
   \longbibliography@sw{%
    ,author="00",editor="1",pages="1",title="0",year="0"%
   }{%
    ,author="48",editor="1",pages="0",title="",year="1"%
   }%
  }%
 }%
 \if@filesw
  \immediate\write\@auxout{\string\citation{apsrev41Control}}%
 \fi
}%
\makeatother

\def\imo{i}
\def\K{{\cal K}}
\def\Order#1{{\cal O}\left(#1\right)}

\begin{document}

\title{Wave and particle probes of a regular T-duality-inspired black hole with gravitational self-energy}

\author{Alexey~Dubinsky}
\email{dubinsky@ukr.net}
\affiliation{University of Seville, Seville, Spain}

\begin{abstract}
Recently it was shown that a non-local T-duality-inspired smearing of the point mass introduces a finite zero-point length, while the regularized Newtonian gravitational self-energy is promoted to an additional source for the spacetime.  The result is a nonsingular black-hole geometry whose ADM mass contains a finite self-energy contribution and whose extremal Planck-scale remnant sector has been proposed as a possible dark-matter component.  We study how this spacetime would affect two familiar physical signals: the ringing of a massive scalar field and the motion of particles and light near the horizon.  The zero-point length smooths the central region and changes the strong-field potential outside the horizon. On the wave side, we find that making the scalar field heavier increases the oscillation frequency and makes the damping weaker, a behavior associated with long-lived ringing.  On the particle side, increasing the zero-point length makes the photon orbit and the innermost stable circular orbit more compact in physical mass units.  The corresponding shadow becomes smaller, the photon-ring frequency becomes larger, and the orbital binding energy increases.  These results show that the same regularizing correction leaves related imprints in wave propagation, black-hole shadows and circular-orbit physics.
\end{abstract}

\maketitle

\section{Introduction}

Regular black holes provide effective geometries in which the central curvature singularity is replaced by a finite core, while the exterior region can remain close to the Schwarzschild solution.  They are useful laboratories for testing how short-distance corrections, non-locality, or effective matter sources can influence observables outside the event horizon.  Even when the horizon position changes only moderately, the effective potentials controlling waves and particles may be modified enough to shift quasinormal frequencies, photon-ring properties, shadows, and circular-orbit observables.  For this reason, regular black holes are often used as phenomenological probes of possible quantum-gravity corrections and of alternative compact-object models \cite{Konoplya:2023ppx,Li:2014fka,MahdavianYekta:2019pol,Al-Badawi:2023lke,Konoplya:2024kih,Huang:2023aet,Lin:2013ofa,Guo:2024jhg,Cai:2021ele,Konoplya:2024hfg,Yang:2021cvh,Mukohyama:2023xyf,Gingrich:2024tuf,Held:2019xde,Konoplya:2025ect,Lopez:2022uie,Pedraza:2021hzw,Al-Badawi:2023lke,Jawad:2020hju,Konoplya:2025hgp,Zhang:2024nny,Flachi:2012nv,Konoplya:2023aph,Jusufi:2020odz,Fernando:2012yw,DuttaRoy:2022ytr,Konoplya:2022hll,Bolokhov:2026kqu,Bolokhov:2026eqf,Skvortsova:2024wly}.

Quasinormal modes are especially well suited to this purpose. They are defined by purely ingoing waves at the event horizon and outgoing waves, or the appropriate massive-field continuation, at spatial infinity.  The resulting complex frequencies are independent of the details of the initial perturbation and characterize the linear response of the background.  The real part fixes the oscillation frequency, while the negative imaginary part fixes the damping time.  General reviews and standard references on black-hole quasinormal modes include Refs.~\cite{Berti:2009kk,Kokkotas:1999bd,Konoplya:2011qq,Nollert:1999ji,Bolokhov:2025rng}.

Particle motion gives an independent but closely related probe.  In a static spherical spacetime the unstable null circular orbit controls the geometrical-optics limit of massless perturbations and determines the critical impact parameter seen as the shadow radius.  The same orbit also provides the eikonal relation between the real and imaginary parts of high-multipole quasinormal modes and, respectively, the orbital frequency and Lyapunov exponent of the photon ring~\cite{Cardoso:2008bp}.  Timelike circular orbits probe the matter-accretion side of the same geometry through the ISCO radius and the binding energy released by a particle moving from rest at infinity to the ISCO.  The shadow and photon-ring literature provides a complementary observational language for these quantities~\cite{Perlick:2021aok}.

Massive perturbations sharpen this comparison between wave and particle probes.  The mass need not be interpreted only as a fundamental scalar mass: effective mass scales can be generated for brane-localized fields by higher-dimensional bulk physics~\cite{Seahra:2004fg}, can enter massive-gravity descriptions of extremely long wavelength gravitational radiation~\cite{Konoplya:2023fmh,NANOGrav:2023hvm}, and can arise for otherwise massless test fields in magnetized black-hole backgrounds~\cite{Konoplya:2008hj,Wu:2015fwa}.  Once such a scale is present, the radial potential approaches the field-mass threshold instead of vanishing at infinity.  The field mass therefore controls the asymptotic plateau, while the geometry still fixes the near-horizon and barrier region.

Several effects then appear that are absent, or much less pronounced, in the massless problem.  A well-known example is the appearance of quasiresonances, in which the damping rate can become arbitrarily small at selected values of the mass parameter~\cite{Ohashi:2004wr,Konoplya:2004wg}.  Long-lived massive modes and related spectral deformations have been found for many spins and in many black-hole or compact-object backgrounds~\cite{Churilova:2020bql,Burikham:2017gdm,Konoplya:2017tvu,Percival:2020skc,Gonzalez:2022upu,Zinhailo:2018ska,Zhidenko:2006rs,Konoplya:2018qov,Aragon:2020teq,Fernandes:2021qvr,Bolokhov:2026dfg,Skvortsova:2026jtx}.  More recent studies have extended this picture to additional regular, deformed and modified-gravity geometries~\cite{Lutfuoglu:2025eik,Lutfuoglu:2025hwh,Skvortsova:2026unq,Bolokhov:2024bke,Skvortsova:2025cah,Lutfuoglu:2025qkt,Lutfuoglu:2025kqp,Konoplya:2007zx,Lutfuoglu:2026fpx,Bolokhov:2026dzn,Lutfuoglu:2025bsf,Skvortsova:2024eqi,Lutfuoglu:2025hjy,Bolokhov:2023bwm,Bolokhov:2023ruj,Bolokhov:2026uol}.  The time-domain signal is also altered: the familiar massless power-law tail is replaced by oscillatory massive tails whose decay depends on the mass scale and on the asymptotic geometry~\cite{Gibbons:2008rs,Dubinsky:2024jqi,Jing:2004zb,Moderski:2001tk,Koyama:2001ee,Koyama:2001qw,Rogatko:2007zz,Gibbons:2008gg,Koyama:2000hj}.  At the same time, the quasiresonant tendency is not guaranteed in every background or every sector~\cite{Zinhailo:2024jzt,Konoplya:2005hr}.  For the self-energy black hole studied below, the massive scalar field therefore tests not only the barrier height, but also how the zero-point-length deformation competes with the massive asymptotic plateau.

The background considered in this work is the neutral regular black hole constructed from a T-duality-inspired zero-point length and a regularized gravitational self-energy~\cite{Jusufi:2025selfenergy}.  For the same metric, Hawking radiation and evaporation of massless fields were analyzed in Ref.~\cite{3164933}, while massless scalar, electromagnetic, and Dirac test-field spectra and excitation factors were obtained in Ref.~\cite{Skvortsova:2026ryl}.  Here we treat the massive scalar ringdown and geodesic motion in a single framework.  The central question is how the zero-point length changes the effective-potential structure outside the horizon and how this change appears in wave and particle observables.

The calculation is naturally expressed in ADM units.  The bare parameter $M$ is used in the metric, but the physical asymptotic mass is $M_{\rm ADM}=M+3\pi M^2/(32l_0)$.  We therefore use $l_0/M_{\rm ADM}$ as the dimensionless deformation parameter and $\hat\mu=\mu_sM_{\rm ADM}$ as the dimensionless scalar mass.  The scalar calculation shows that increasing $\hat\mu$ raises ${\rm Re}(M_{\rm ADM}\omega)$ and reduces the damping rate $-M_{\rm ADM}{\rm Im}(\omega)$ along each reliable WKB branch.  The geodesic calculation shows that increasing $l_0/M_{\rm ADM}$ shifts the photon sphere and ISCO inward, reduces the shadow radius, and increases the efficiency of the innermost stable orbit.

The paper is organized as follows.  In Sec.~\ref{sec:spacetime} we review the regular self-energy black-hole metric and the ADM normalization.  In Sec.~\ref{sec:scalar-equation} we derive the massive Klein--Gordon radial equation, its effective potential and the quasinormal-mode boundary conditions used in the calculation.  Section~\ref{sec:geodesics} derives the particle-motion quantities.  Section~\ref{sec:numerics} summarizes the WKB--Pad\'e calculation and the numerical extraction of the geodesic observables.  Section~\ref{sec:qnm-adm} presents the massive scalar spectrum, and Sec.~\ref{sec:geodesic-results} presents the photon-ring, ISCO and shadow results.  Section~\ref{sec:conclusions} summarizes the combined picture.

\section{Regular self-energy black hole}\label{sec:spacetime}

The geometry used in this work is an effective classical spacetime motivated by a non-local gravitational model inspired by T-duality~\cite{Jusufi:2025selfenergy}.  The central assumption is that the point-particle source of the Schwarzschild solution is replaced by a smooth distribution with a finite zero-point length $l_0$.  In the notation of Ref.~\cite{Jusufi:2025selfenergy}, the regularized bare density and its Newtonian potential may be written as
\begin{align*}
 \rho_{\rm b}(r)&=\frac{3Ml_0^2}{4\pi(r^2+l_0^2)^{5/2}},\\
 V_G(r)&=-\frac{M}{\sqrt{r^2+l_0^2}} .
\end{align*}
Thus $l_0$ acts as a short-distance cutoff in the source and prevents the matter density from becoming singular at the origin.  In the limit where the non-local scale is removed, the exterior geometry reduces to the Schwarzschild one after the mass is expressed in terms of the ADM mass.

The same scale also regularizes the Newtonian gravitational self-energy.  In geometrized units the corresponding effective density is
\begin{align*}
 \rho_{\rm GSE}(r)
 &=\frac{1}{8\pi}\left(\frac{dV_G}{dr}\right)^2
 =\frac{M^2r^2}{8\pi(r^2+l_0^2)^3},\\
 m(r)&=4\pi\int_0^r
 \left[\rho_{\rm b}(r')+\rho_{\rm GSE}(r')\right]r'^2dr' .
\end{align*}
Instead of treating the self-energy as a divergent quantity to be discarded, the construction includes this finite contribution as part of the effective energy density sourcing the metric.  The result is an electrically neutral regular black hole: the parameter $l_0$ plays the role of the regularizing scale, while no electromagnetic charge is required.  The present paper does not solve the underlying non-local field equations again; it takes the resulting static and spherically symmetric metric as a background and studies test scalar waves and geodesic motion on it.

The line element is written in Schwarzschild-like coordinates as
\begin{eqnarray}\label{metric}
 ds^2&=&-f(r)dt^2+\frac{dr^2}{f(r)}+r^2(d\theta^2+\sin^2\theta d\phi^2),\nonumber\\
 f(r)&=&1-\frac{2m(r)}{r} .
\end{eqnarray}
The mass function follows from the sum of the regular bare density and the gravitational-self-energy density.  For the black hole solution of Ref.~\cite{Jusufi:2025selfenergy} it is
\begin{widetext}
\begin{equation}\label{mass-function}
 m(r)=\frac{Mr^3}{\left(r^2+l_0^2\right)^{3/2}}
 +\frac{3M^2}{16l_0}\tan^{-1}\left(\frac{r}{l_0}\right)
 -\frac{3M^2l_0^2r}{16\left(r^2+l_0^2\right)^2}
 -\frac{5M^2r^3}{16\left(r^2+l_0^2\right)^2} .
\end{equation}
\end{widetext}
Equivalently, we have
\begin{widetext}
\begin{eqnarray}\label{metric-function}
 f(r)&=&1-\frac{2 M r^2}{\left(l_0^2+r^2\right)^{3/2}}
 -\frac{3M^2}{8l_0 r}\tan^{-1}\left(\frac{r}{l_0}\right)
 +\frac{3l_0^2M^2}{8\left(l_0^2+r^2\right)^2}
 +\frac{5M^2r^2}{8\left(l_0^2+r^2\right)^2} .
\end{eqnarray}
\end{widetext}
Here $M$ is the bare mass parameter and $l_0$ is the zero-point length.  The ADM mass is obtained from the asymptotic value of $m(r)$,
\begin{equation}\label{ADMmass}
 M_{\rm ADM}=M+\frac{3\pi M^2}{32l_0} .
\end{equation}
In the numerical work it is useful to set $M=1$ internally and convert the final results to ADM variables.  The relevant dimensionless deformation parameter is therefore
\begin{equation}\label{alpha-def}
 \alpha=\frac{l_0}{M_{\rm ADM}} .
\end{equation}
The scalar-field mass is similarly measured by
\begin{equation}\label{mu-def}
 \hat\mu=\mu_s M_{\rm ADM} .
\end{equation}

The event horizon $r_+$ is the largest positive root of $f(r)=0$.  The tortoise coordinate is defined by
\begin{equation}\label{tortoise}
 \frac{dr_*}{dr}=\frac{1}{f(r)} ,
\end{equation}
so that $r_*\to-\infty$ at the event horizon.  At large radius the geometry is asymptotically flat and
\begin{equation}
 f(r)=1-\frac{2M_{\rm ADM}}{r}+\frac{M^2}{r^2}+\Order{r^{-3}} .
\end{equation}

\section{Massive scalar wave equation}\label{sec:scalar-equation}

We consider a minimally coupled scalar field of mass $\mu_s$ obeying
\begin{equation}\label{KG}
 \frac{1}{\sqrt{-g}}\partial_\mu\left(\sqrt{-g}g^{\mu\nu}\partial_\nu\Phi\right)-\mu_s^2\Phi=0 .
\end{equation}
After separating variables as {(see, for instance \cite{Carter:1968ks,Konoplya:2018arm})}
\begin{equation}\label{separation}
 \Phi(t,r,\theta,\phi)=\frac{\psi_{\ell}(r)}{r}Y_{\ell m}(\theta,\phi)e^{-\imo\omega t} ,
\end{equation}
the radial equation takes the Schr\"odinger-like form
\begin{equation}\label{wave-equation}
 \frac{d^2\psi_\ell}{dr_*^2}+\left[\omega^2-V_\ell^{(s)}(r)\right]\psi_\ell=0 .
\end{equation}
The massive scalar effective potential is
\begin{equation}\label{scalar-potential}
 V_\ell^{(s)}(r)=f(r)\left[\frac{\ell(\ell+1)}{r^2}+\frac{f'(r)}{r}+\mu_s^2\right] .
\end{equation}
In the massless limit this reduces to the usual scalar-field potential for a static spherically symmetric black hole.  The term proportional to $\mu_s^2$ changes the asymptotic value of the potential,
\begin{equation}\label{asymptotic-potential}
 V_\ell^{(s)}(r\to\infty)\to\mu_s^2,
\end{equation}
and is responsible for the main qualitative differences between massive and massless ringdown.

In ADM units, the potential can be written as
\begin{equation}\label{adm-potential}
 M_{\rm ADM}^2V_\ell^{(s)}=F(\bar r)\left[\frac{\ell(\ell+1)}{\bar r^2}+\frac{F'(\bar r)}{\bar r}+\hat\mu^2\right] ,
\end{equation}
where $\bar r=r/M_{\rm ADM}$, $F(\bar r)=f(r)$, and the prime now denotes differentiation with respect to $\bar r$.  This form makes clear that the spectrum is governed by $\alpha=l_0/M_{\rm ADM}$ and $\hat\mu=\mu_sM_{\rm ADM}$.

The quasinormal-mode boundary conditions used below are imposed on the same radial equation.  Near the event horizon the potential vanishes and the physical solution is purely ingoing,
\begin{equation}\label{horizon-bc}
 \psi_\ell\sim e^{-\imo\omega r_*},\qquad r\to r_+ .
\end{equation}
At infinity the massive field behaves differently from a massless field.  Defining
\begin{equation}\label{chi-def}
 \chi=\sqrt{\omega^2-\mu_s^2} ,
\end{equation}
the quasinormal mode satisfies the outgoing-wave condition
\begin{equation}\label{qnm-infinity-bc}
 \psi_\ell\sim e^{+\imo\chi r_*},\qquad r\to\infty,
\end{equation}
with the branch of $\chi$ chosen by analytic continuation from the real-frequency scattering problem.

\section{Particle motion and optical observables}\label{sec:geodesics}

The geodesic problem is governed by the same metric function $f(r)$ that enters the wave potential.  Let $u^\mu=dx^\mu/d\sigma$, where $\sigma$ is proper time for a timelike worldline and an affine parameter for a null ray.  The normalization is
\begin{equation}\label{geodesic-normalization}
 g_{\mu\nu}u^\mu u^\nu=-\varepsilon,
\end{equation}
where $\varepsilon=0$ for null geodesics and $\varepsilon=1$ for timelike geodesics with unit rest mass.
Spherical symmetry allows the orbit to be placed in the equatorial plane, $\theta=\pi/2$.  The Killing vectors $\partial_t$ and $\partial_\phi$ then give the conserved quantities
\begin{equation}\label{energy-angular-momentum}
 E=f(r)\dot t,
 \qquad
 L=r^2\dot\phi,
\end{equation}
where the dot denotes $d/d\sigma$.  For timelike geodesics $E$ and $L$ are the specific energy and specific angular momentum.  For null geodesics their common affine rescaling is arbitrary, and the physical quantity is the impact parameter $b=L/E$.

Using Eq.~(\ref{geodesic-normalization}), the radial equation can be written as
\begin{equation}\label{geodesic-potential}
 \dot r^2+V_{\rm geo}(r)=E^2,
 \qquad
 V_{\rm geo}(r)=f(r)\left(\varepsilon+\frac{L^2}{r^2}\right).
\end{equation}
Thus circular orbits are stationary points of $V_{\rm geo}$ at fixed $L$, with $V_{\rm geo}=E^2$.

For null geodesics, Eq.~(\ref{geodesic-potential}) may be divided by $E^2$ and written in terms of the optical potential
\begin{equation}\label{optical-potential}
 U_{\rm opt}(r)=\frac{f(r)}{r^2},
 \qquad
 \frac{\dot r^2}{E^2}=1-b^2U_{\rm opt}(r).
\end{equation}
The photon-ring radius $r_{\rm ph}$ is therefore an extremum of $U_{\rm opt}$.  The radius used below is the outer unstable solution outside the event horizon, determined by
\begin{equation}\label{photon-condition}
 r_{\rm ph}f'(r_{\rm ph})-2f(r_{\rm ph})=0,
\end{equation}
where the prime denotes $d/dr$.  The corresponding critical impact parameter is
\begin{equation}\label{critical-impact-parameter}
 b_{\rm ph}=\frac{L}{E}=\frac{r_{\rm ph}}{\sqrt{f(r_{\rm ph})}} .
\end{equation}
For an observer at infinity in this asymptotically flat spacetime, $b_{\rm ph}$ is the shadow radius.  The coordinate angular frequency, Lyapunov exponent and shadow radius are then
\begin{eqnarray}
 \Omega_{\rm ph}&=&\left.\frac{d\phi}{dt}\right|_{r_{\rm ph}}
 =\frac{\sqrt{f(r_{\rm ph})}}{r_{\rm ph}},\label{omega-ph}\\
 \lambda&=&\left[\frac{f(r_{\rm ph})\left(2f(r_{\rm ph})-r_{\rm ph}^2f''(r_{\rm ph})\right)}{2r_{\rm ph}^2}\right]^{1/2},\label{lambda-ph}\\
 R_{\rm sh}&=&b_{\rm ph}=\frac{r_{\rm ph}}{\sqrt{f(r_{\rm ph})}}=\frac{1}{\Omega_{\rm ph}} .\label{shadow-radius}
\end{eqnarray}
The Lyapunov exponent follows from the second-order expansion of the null radial equation about the maximum of $U_{\rm opt}$ and measures the instability time scale in coordinate time.  In the eikonal limit of massless fields it enters the standard correspondence
\begin{equation}\label{eikonal-qnm}
 \omega_{\ell n}\simeq \ell\Omega_{\rm ph}-\imo\left(n+\frac{1}{2}\right)\lambda .
\end{equation}
The massive scalar modes studied below are not restricted to this eikonal regime, but Eq.~(\ref{eikonal-qnm}) explains why photon-ring quantities and wave spectra are complementary probes of the same effective-potential deformation.
Notice, however, that this correspondence between null geodesics and quasinormal modes is not universal and may break down in some cases, especially when the eikonal limit of  an alternative theory of gravity has a non-standard form \cite{Konoplya:2022gjp,Bolokhov:2023dxq,Konoplya:2017ymp}. In our case, the centrifugal barrier has the usual form $f(r) \ell (\ell+1) r^{-2}$ and the correspondence is fulfilled.

For timelike circular geodesics, the conditions $V_{\rm geo}=E^2$ and $V'_{\rm geo}=0$ give
\begin{eqnarray}
 E^2(r)&=&\frac{2f(r)^2}{2f(r)-rf'(r)},\label{timelike-energy}\\
 L^2(r)&=&\frac{r^3f'(r)}{2f(r)-rf'(r)},\label{timelike-angular-momentum}\\
 \Omega_{\rm K}(r)&=&\sqrt{\frac{f'(r)}{2r}} .\label{kepler-frequency}
\end{eqnarray}
Here $\Omega_{\rm K}=d\phi/dt$ is the coordinate Keplerian frequency and the circular-orbit branch requires $2f-rf'>0$.  Stability is controlled by the second radial derivative of $V_{\rm geo}$ at fixed $L$.  The innermost stable circular orbit is the marginal point, continuously connected to the Schwarzschild ISCO, where this second derivative vanishes.  Equivalently, it satisfies
\begin{equation}\label{isco-condition}
\begin{aligned}
0={}&f(r_{\rm ISCO})f''(r_{\rm ISCO})
 -2f'(r_{\rm ISCO})^2\\
&+\frac{3f(r_{\rm ISCO})f'(r_{\rm ISCO})}{r_{\rm ISCO}} .
\end{aligned}
\end{equation}
The specific binding energy at the ISCO is
\begin{equation}\label{binding-energy}
 E_b=1-E(r_{\rm ISCO}) .
\end{equation}
Here $E(r_{\rm ISCO})$ is the positive square root of Eq.~(\ref{timelike-energy}).  Because a particle at rest at infinity has specific energy $E=1$, this quantity is the usual thin-disk efficiency proxy: larger $E_b$ means that more rest-mass energy can be released before matter crosses the horizon.

\section{Numerical strategy}\label{sec:numerics}

The numerical frequencies are obtained with the WKB expansion around the maximum of the effective potential.  Let $r_0$ be the position of the peak of $V_\ell^{(s)}(r)$ in the tortoise coordinate and let $V_j$ denote the $j$th derivative of the potential with respect to $r_*$ at $r_0$.  In the notation of Ref.~\cite{Konoplya:2019hlu}, the WKB condition is
\begin{eqnarray}\label{WKBformula}
\omega^2&=&V_0+A_2(\K^2)+A_4(\K^2)+A_6(\K^2)+\cdots\nonumber\\
&&-\imo\K\sqrt{-2V_2}\left[1+A_3(\K^2)+A_5(\K^2)+\cdots\right],
\end{eqnarray}
where $\K=n+1/2$ and $n$ is the overtone number.  The functions $A_j$ are built from higher derivatives of the potential at the peak.  The higher-order terms were derived in Refs.~\cite{Iyer:1986np,Konoplya:2003ii,Matyjasek:2019eeu}, and Pad\'e resummation is used here to improve the behavior of the high-order series \cite{Matyjasek:2017psv,Konoplya:2026rjh}. The high-order WKB method has been extensively used for finding quasinormal modes and grey-body factors of black holes \cite{Lutfuoglu:2026xlo,Skvortsova:2024msa,Kokkotas:2010zd,Bolokhov:2025aqy,Bolokhov:2025lnt,Konoplya:2010vz,Breton:2017hwe,Fernando:2016ftj,Konoplya:2023moy,Malik:2025erb,Konoplya:2002wt,Bolokhov:2025egl,Konoplya:2009hv,Pathrikar:2025gzu,Wongjun:2019ydo,Guo:2020caw,Malik:2026lfj,Malik:2025qnr,Konoplya:2019ppy,Eniceicu:2019npi,Malik:2025czt,Ishihara:2008re,Skvortsova:2023zmj,Lutfuoglu:2026zxj}.

We compute the fundamental mode $n=0$ for $\ell=0,1,2$ and for three representative zero-point lengths, $l_0=0.1$, $0.5$, and $0.8$, with the bare mass set to $M=1$ during the WKB calculation.  For each value of $l_0$ and $\ell$ the scalar mass $\mu_s$ is increased until the potential no longer has a suitable single barrier or until the WKB sequence becomes unstable.  The tables display only entries for which a peak is present and for which the relative difference between the 16th- and 14th-order Pad\'e-resummed WKB values does not exceed one percent.  The relative difference is evaluated as
\begin{equation}\label{difference-definition}
 \Delta=100\,\frac{|\omega_{16}-\omega_{14}|}{|\omega_{16}|}\, .
\end{equation}
This criterion removes the data where the massive term has already pushed the system outside the range in which the barrier WKB approximation is reliable.

The frequency convention is $e^{-\imo\omega t}$.  Stable quasinormal modes therefore have ${\rm Im}(\omega)<0$.  In the discussion below, we use the positive damping rate
\begin{equation}\label{gamma-definition}
 \Gamma=-{\rm Im}(\Omega),\qquad \Omega=M_{\rm ADM}\omega,
\end{equation}
where $\Omega$ is the ADM-scaled frequency.  The scalar mass is similarly written as $\hat\mu=\mu_sM_{\rm ADM}$.  These variables allow spectra with different values of $l_0$ to be compared at fixed physical mass.

The geodesic quantities are obtained by solving Eqs.~(\ref{photon-condition}) and (\ref{isco-condition}) outside the event horizon.  For the photon ring we use the outer unstable extremum of $U_{\rm opt}$, and for the ISCO we use the marginal point of the stable timelike circular-orbit branch.  All reported radii are divided by $M_{\rm ADM}$, whereas frequencies and Lyapunov exponents are multiplied by $M_{\rm ADM}$.  The same normalization is used for the shadow radius.  This makes the geodesic table directly comparable with the ADM-scaled wave frequencies.

\section{Quasinormal modes in ADM units}\label{sec:qnm-adm}

Tables~\ref{tab:qnm-l001}, \ref{tab:qnm-l005}, and \ref{tab:qnm-l008} list the fundamental massive scalar modes in the working normalization $M=1$.  The table headings use $\mu_s$ for the scalar-field mass in the same units.  Each table combines the $\ell=0,1,2$ sectors for one fixed value of $l_0$.  Blank entries only indicate that the reliable range of $\mu_s$ is shorter for that multipole; rows with no potential peak or with $\Delta>1\%$ are not shown.

\begin{table*}
\centering
\scriptsize
\setlength{\tabcolsep}{2.0pt}
\renewcommand{\arraystretch}{0.9}
\resizebox{\textwidth}{!}{%
\begin{tabular}{c c c c c c c c c c c c}
\hline
\multicolumn{4}{c}{$\ell=0$} & \multicolumn{4}{c}{$\ell=1$} & \multicolumn{4}{c}{$\ell=2$}\\
\hline
$\mu_s$ & WKB16 & WKB14 & diff. & $\mu_s$ & WKB16 & WKB14 & diff. & $\mu_s$ & WKB16 & WKB14 & diff.\\
\hline
$0$ & $0.028328-0.026669 i$ & $0.028305-0.026582 i$ & $0.230\%$ & $0$ & $0.075080-0.024835 i$ & $0.075080-0.024835 i$ & $0\%$ & $0$ & $0.123948-0.024609 i$ & $0.123948-0.024609 i$ & $0\%$\\
$0.01$ & $0.028419-0.026354 i$ & $0.028392-0.026256 i$ & $0.264\%$ & $0.01$ & $0.075254-0.024732 i$ & $0.075254-0.024732 i$ & $0\%$ & $0.01$ & $0.124071-0.024568 i$ & $0.124071-0.024568 i$ & $0\%$\\
$0.02$ & $0.028692-0.025449 i$ & $0.028562-0.025231 i$ & $0.662\%$ & $0.02$ & $0.075776-0.024420 i$ & $0.075776-0.024420 i$ & $0\%$ & $0.02$ & $0.124438-0.024442 i$ & $0.124438-0.024442 i$ & $0\%$\\
$0.03$ & $0.029111-0.023829 i$ & $0.029323-0.023853 i$ & $0.568\%$ & $0.03$ & $0.076647-0.023895 i$ & $0.076647-0.023895 i$ & $0\%$ & $0.03$ & $0.125051-0.024233 i$ & $0.125051-0.024233 i$ & $0\%$\\
$0.04$ & $0.029564-0.022038 i$ & $0.029582-0.021851 i$ & $0.510\%$ & $0.04$ & $0.077871-0.023149 i$ & $0.077871-0.023149 i$ & $0\%$ & $0.04$ & $0.125912-0.023939 i$ & $0.125912-0.023939 i$ & $0\%$\\
 &  &  &  & $0.05$ & $0.079452-0.022171 i$ & $0.079452-0.022171 i$ & $0\%$ & $0.05$ & $0.127021-0.023557 i$ & $0.127021-0.023557 i$ & $0\%$\\
 &  &  &  & $0.06$ & $0.081393-0.020945 i$ & $0.081393-0.020945 i$ & $0\%$ & $0.06$ & $0.128381-0.023088 i$ & $0.128381-0.023088 i$ & $0\%$\\
 &  &  &  & $0.07$ & $0.083698-0.019447 i$ & $0.083698-0.019446 i$ & $0.0001\%$ & $0.07$ & $0.129996-0.022527 i$ & $0.129996-0.022527 i$ & $0\%$\\
 &  &  &  & $0.08$ & $0.086363-0.017647 i$ & $0.086363-0.017647 i$ & $0\%$ & $0.08$ & $0.131869-0.021872 i$ & $0.131869-0.021872 i$ & $0\%$\\
 &  &  &  & $0.09$ & $0.089377-0.015509 i$ & $0.089378-0.015509 i$ & $0.0008\%$ & $0.09$ & $0.134004-0.021118 i$ & $0.134004-0.021118 i$ & $0\%$\\
 &  &  &  & $0.1$ & $0.092714-0.012995 i$ & $0.092721-0.013004 i$ & $0.0117\%$ & $0.1$ & $0.136406-0.020261 i$ & $0.136406-0.020261 i$ & $0\%$\\
 &  &  &  & $0.11$ & $0.096285-0.010096 i$ & $0.096281-0.010097 i$ & $0.0042\%$ & $0.11$ & $0.139082-0.019292 i$ & $0.139082-0.019292 i$ & $0\%$\\
 &  &  &  &  &  &  &  & $0.12$ & $0.142037-0.018204 i$ & $0.142037-0.018204 i$ & $0\%$\\
 &  &  &  &  &  &  &  & $0.13$ & $0.145278-0.016985 i$ & $0.145278-0.016985 i$ & $0\%$\\
 &  &  &  &  &  &  &  & $0.14$ & $0.148812-0.015619 i$ & $0.148812-0.015619 i$ & $0\%$\\
 &  &  &  &  &  &  &  & $0.15$ & $0.152644-0.014089 i$ & $0.152644-0.014089 i$ & $0\%$\\
 &  &  &  &  &  &  &  & $0.16$ & $0.156775-0.012371 i$ & $0.156775-0.012371 i$ & $0\%$\\
 &  &  &  &  &  &  &  & $0.17$ & $0.161200-0.010438 i$ & $0.161199-0.010438 i$ & $0.0007\%$\\
 &  &  &  &  &  &  &  & $0.18$ & $0.165894-0.008255 i$ & $0.165878-0.008260 i$ & $0.0105\%$\\
\hline
\end{tabular}%
}
\caption{Massive scalar quasinormal modes for fixed $l_0=0.1$ ($M=1$).  The table combines the $\ell=0,1,2$ data obtained from the 16th- and 14th-order Pad\'e-resummed WKB approximants.  The last column in each block gives the relative difference in percent.}
\label{tab:qnm-l001}
\end{table*}

\begin{table*}
\centering
\scriptsize
\setlength{\tabcolsep}{2.0pt}
\renewcommand{\arraystretch}{0.9}
\resizebox{\textwidth}{!}{%
\begin{tabular}{c c c c c c c c c c c c}
\hline
\multicolumn{4}{c}{$\ell=0$} & \multicolumn{4}{c}{$\ell=1$} & \multicolumn{4}{c}{$\ell=2$}\\
\hline
$\mu_s$ & WKB16 & WKB14 & diff. & $\mu_s$ & WKB16 & WKB14 & diff. & $\mu_s$ & WKB16 & WKB14 & diff.\\
\hline
$0$ & $0.077044-0.065711 i$ & $0.076945-0.065479 i$ & $0.249\%$ & $0$ & $0.203410-0.061704 i$ & $0.203410-0.061704 i$ & $0\%$ & $0$ & $0.335546-0.061229 i$ & $0.335546-0.061229 i$ & $0\%$\\
$0.01$ & $0.077087-0.065600 i$ & $0.076986-0.065365 i$ & $0.253\%$ & $0.01$ & $0.203471-0.061671 i$ & $0.203471-0.061671 i$ & $0\%$ & $0.02$ & $0.335715-0.061175 i$ & $0.335715-0.061175 i$ & $0\%$\\
$0.02$ & $0.077213-0.065269 i$ & $0.077106-0.065024 i$ & $0.265\%$ & $0.02$ & $0.203654-0.061571 i$ & $0.203654-0.061571 i$ & $0\%$ & $0.04$ & $0.336223-0.061015 i$ & $0.336223-0.061015 i$ & $0\%$\\
$0.03$ & $0.077420-0.064721 i$ & $0.077297-0.064454 i$ & $0.291\%$ & $0.03$ & $0.203960-0.061404 i$ & $0.203960-0.061404 i$ & $0\%$ & $0.06$ & $0.337071-0.060748 i$ & $0.337071-0.060748 i$ & $0\%$\\
$0.04$ & $0.077706-0.063959 i$ & $0.077547-0.063650 i$ & $0.345\%$ & $0.04$ & $0.204388-0.061169 i$ & $0.204388-0.061169 i$ & $0\%$ & $0.08$ & $0.338260-0.060373 i$ & $0.338260-0.060373 i$ & $0\%$\\
$0.05$ & $0.078079-0.062985 i$ & $0.077824-0.062585 i$ & $0.473\%$ & $0.05$ & $0.204939-0.060867 i$ & $0.204939-0.060867 i$ & $0\%$ & $0.1$ & $0.339790-0.059888 i$ & $0.339790-0.059888 i$ & $0\%$\\
$0.06$ & $0.078537-0.061757 i$ & $0.078115-0.060953 i$ & $0.909\%$ & $0.06$ & $0.205613-0.060496 i$ & $0.205613-0.060496 i$ & $0\%$ & $0.12$ & $0.341665-0.059292 i$ & $0.341665-0.059292 i$ & $0\%$\\
$0.07$ & $0.078972-0.060265 i$ & $0.079525-0.059951 i$ & $0.641\%$ & $0.07$ & $0.206411-0.060055 i$ & $0.206411-0.060055 i$ & $0\%$ & $0.14$ & $0.343887-0.058583 i$ & $0.343887-0.058583 i$ & $0\%$\\
$0.08$ & $0.079404-0.058590 i$ & $0.079723-0.058574 i$ & $0.324\%$ & $0.08$ & $0.207332-0.059544 i$ & $0.207332-0.059544 i$ & $0\%$ & $0.16$ & $0.346459-0.057757 i$ & $0.346459-0.057757 i$ & $0\%$\\
$0.09$ & $0.080110-0.056952 i$ & $0.080111-0.056952 i$ & $0.0003\%$ & $0.09$ & $0.208377-0.058960 i$ & $0.208377-0.058960 i$ & $0\%$ & $0.18$ & $0.349385-0.056811 i$ & $0.349385-0.056811 i$ & $0\%$\\
$0.1$ & $0.080507-0.055327 i$ & $0.080620-0.054837 i$ & $0.515\%$ & $0.1$ & $0.209547-0.058303 i$ & $0.209547-0.058303 i$ & $0\%$ & $0.2$ & $0.352668-0.055742 i$ & $0.352668-0.055742 i$ & $0\%$\\
$0.12$ & $0.080445-0.051982 i$ & $0.080440-0.051989 i$ & $0.0080\%$ & $0.11$ & $0.210842-0.057571 i$ & $0.210842-0.057571 i$ & $0\%$ & $0.22$ & $0.356313-0.054543 i$ & $0.356313-0.054543 i$ & $0\%$\\
$0.14$ & $0.079733-0.046938 i$ & $0.079658-0.046988 i$ & $0.0968\%$ & $0.12$ & $0.212263-0.056762 i$ & $0.212263-0.056762 i$ & $0\%$ & $0.24$ & $0.360327-0.053210 i$ & $0.360327-0.053210 i$ & $0\%$\\
 &  &  &  & $0.13$ & $0.213810-0.055873 i$ & $0.213810-0.055873 i$ & $0\%$ & $0.26$ & $0.364714-0.051736 i$ & $0.364714-0.051736 i$ & $0\%$\\
 &  &  &  & $0.14$ & $0.215483-0.054902 i$ & $0.215484-0.054902 i$ & $0\%$ & $0.28$ & $0.369481-0.050111 i$ & $0.369481-0.050111 i$ & $0\%$\\
 &  &  &  & $0.15$ & $0.217284-0.053846 i$ & $0.217284-0.053846 i$ & $0\%$ & $0.3$ & $0.374636-0.048327 i$ & $0.374636-0.048327 i$ & $0\%$\\
 &  &  &  & $0.16$ & $0.219212-0.052701 i$ & $0.219212-0.052701 i$ & $<10^{-4}\%$ & $0.32$ & $0.380186-0.046370 i$ & $0.380186-0.046370 i$ & $0\%$\\
 &  &  &  & $0.17$ & $0.221267-0.051465 i$ & $0.221266-0.051465 i$ & $<10^{-4}\%$ & $0.34$ & $0.386139-0.044227 i$ & $0.386139-0.044227 i$ & $0\%$\\
 &  &  &  & $0.18$ & $0.223449-0.050132 i$ & $0.223449-0.050132 i$ & $0\%$ & $0.36$ & $0.392502-0.041879 i$ & $0.392502-0.041879 i$ & $0\%$\\
 &  &  &  & $0.19$ & $0.225757-0.048699 i$ & $0.225758-0.048699 i$ & $<10^{-4}\%$ & $0.38$ & $0.399282-0.039305 i$ & $0.399282-0.039305 i$ & $0\%$\\
 &  &  &  & $0.2$ & $0.228192-0.047160 i$ & $0.228192-0.047160 i$ & $<10^{-4}\%$ & $0.4$ & $0.406483-0.036479 i$ & $0.406483-0.036479 i$ & $0\%$\\
 &  &  &  & $0.21$ & $0.230750-0.045510 i$ & $0.230751-0.045510 i$ & $<10^{-4}\%$ & $0.42$ & $0.414105-0.033372 i$ & $0.414105-0.033372 i$ & $0.00004\%$\\
 &  &  &  & $0.22$ & $0.233431-0.043744 i$ & $0.233431-0.043744 i$ & $0\%$ & $0.44$ & $0.422143-0.029951 i$ & $0.422142-0.029951 i$ & $0.00003\%$\\
 &  &  &  & $0.23$ & $0.236230-0.041857 i$ & $0.236231-0.041857 i$ & $<10^{-4}\%$ & $0.46$ & $0.430579-0.026182 i$ & $0.430571-0.026185 i$ & $0.00192\%$\\
 &  &  &  & $0.24$ & $0.239144-0.039843 i$ & $0.239147-0.039842 i$ & $0.0011\%$ & $0.48$ & $0.439348-0.022031 i$ & $0.439304-0.022073 i$ & $0.0137\%$\\
 &  &  &  & $0.25$ & $0.242169-0.037698 i$ & $0.242168-0.037692 i$ & $0.0024\%$ & $0.5$ & $0.456874-0.009601 i$ & $0.457382-0.008279 i$ & $0.310\%$\\
 &  &  &  & $0.26$ & $0.245305-0.035405 i$ & $0.245293-0.035417 i$ & $0.0069\%$ &  &  &  & \\
 &  &  &  & $0.27$ & $0.248516-0.032996 i$ & $0.248504-0.033030 i$ & $0.0143\%$ &  &  &  & \\
 &  &  &  & $0.28$ & $0.251802-0.030457 i$ & $0.251802-0.030457 i$ & $0\%$ &  &  &  & \\
 &  &  &  & $0.29$ & $0.255200-0.027733 i$ & $0.255181-0.027731 i$ & $0.0074\%$ &  &  &  & \\
 &  &  &  & $0.3$ & $0.258420-0.025106 i$ & $0.258411-0.025438 i$ & $0.128\%$ &  &  &  & \\
 &  &  &  & $0.31$ & $0.269132-0.022662 i$ & $0.270656-0.021191 i$ & $0.784\%$ &  &  &  & \\
\hline
\end{tabular}%
}
\caption{Massive scalar quasinormal modes for fixed $l_0=0.5$ ($M=1$).  The table combines the $\ell=0,1,2$ data obtained from the 16th- and 14th-order Pad\'e-resummed WKB approximants.  The last column in each block gives the relative difference in percent.}
\label{tab:qnm-l005}
\end{table*}

\begin{table*}
\centering
\scriptsize
\setlength{\tabcolsep}{2.0pt}
\renewcommand{\arraystretch}{0.9}
\resizebox{\textwidth}{!}{%
\begin{tabular}{c c c c c c c c c c c c}
\hline
\multicolumn{4}{c}{$\ell=0$} & \multicolumn{4}{c}{$\ell=1$} & \multicolumn{4}{c}{$\ell=2$}\\
\hline
$\mu_s$ & WKB16 & WKB14 & diff. & $\mu_s$ & WKB16 & WKB14 & diff. & $\mu_s$ & WKB16 & WKB14 & diff.\\
\hline
$0$ & $0.096551-0.066476 i$ & $0.096205-0.066360 i$ & $0.311\%$ & $0$ & $0.267228-0.063147 i$ & $0.267228-0.063147 i$ & $0\%$ & $0$ & $0.442117-0.062743 i$ & $0.442117-0.062743 i$ & $0\%$\\
$0.01$ & $0.096608-0.066405 i$ & $0.096258-0.066287 i$ & $0.315\%$ & $0.02$ & $0.267395-0.063084 i$ & $0.267395-0.063084 i$ & $0\%$ & $0.02$ & $0.442226-0.062719 i$ & $0.442226-0.062719 i$ & $0\%$\\
$0.02$ & $0.096780-0.066193 i$ & $0.096416-0.066067 i$ & $0.329\%$ & $0.04$ & $0.267896-0.062893 i$ & $0.267896-0.062893 i$ & $0\%$ & $0.04$ & $0.442553-0.062645 i$ & $0.442553-0.062645 i$ & $0\%$\\
$0.03$ & $0.097064-0.065843 i$ & $0.096674-0.065700 i$ & $0.354\%$ & $0.06$ & $0.268733-0.062572 i$ & $0.268733-0.062572 i$ & $0\%$ & $0.06$ & $0.443099-0.062523 i$ & $0.443099-0.062523 i$ & $0\%$\\
$0.04$ & $0.097453-0.065371 i$ & $0.097029-0.065189 i$ & $0.393\%$ & $0.08$ & $0.269907-0.062118 i$ & $0.269907-0.062118 i$ & $0\%$ & $0.08$ & $0.443863-0.062351 i$ & $0.443863-0.062351 i$ & $0\%$\\
$0.05$ & $0.097906-0.064818 i$ & $0.097474-0.064538 i$ & $0.439\%$ & $0.1$ & $0.271419-0.061525 i$ & $0.271419-0.061525 i$ & $0\%$ & $0.1$ & $0.444847-0.062129 i$ & $0.444847-0.062129 i$ & $0\%$\\
$0.06$ & $0.098197-0.064098 i$ & $0.098007-0.063762 i$ & $0.329\%$ & $0.12$ & $0.273273-0.060786 i$ & $0.273273-0.060786 i$ & $0\%$ & $0.12$ & $0.446053-0.061854 i$ & $0.446053-0.061854 i$ & $0\%$\\
$0.07$ & $0.098718-0.062850 i$ & $0.098714-0.062844 i$ & $0.0065\%$ & $0.14$ & $0.275472-0.059893 i$ & $0.275472-0.059893 i$ & $0\%$ & $0.14$ & $0.447482-0.061528 i$ & $0.447482-0.061528 i$ & $0\%$\\
$0.08$ & $0.099484-0.061635 i$ & $0.099348-0.061631 i$ & $0.117\%$ & $0.16$ & $0.278019-0.058833 i$ & $0.278019-0.058833 i$ & $<10^{-4}\%$ & $0.16$ & $0.449135-0.061146 i$ & $0.449135-0.061146 i$ & $0\%$\\
$0.09$ & $0.100238-0.060355 i$ & $0.099973-0.060456 i$ & $0.242\%$ & $0.18$ & $0.280917-0.057593 i$ & $0.280917-0.057593 i$ & $<10^{-4}\%$ & $0.18$ & $0.451014-0.060709 i$ & $0.451014-0.060709 i$ & $0\%$\\
$0.1$ & $0.100968-0.058981 i$ & $0.100639-0.059342 i$ & $0.418\%$ & $0.2$ & $0.284169-0.056156 i$ & $0.284169-0.056156 i$ & $0.00006\%$ & $0.2$ & $0.453123-0.060212 i$ & $0.453123-0.060212 i$ & $0\%$\\
$0.11$ & $0.101672-0.057530 i$ & $0.101938-0.058506 i$ & $0.866\%$ & $0.22$ & $0.287777-0.054502 i$ & $0.287777-0.054502 i$ & $<10^{-4}\%$ & $0.22$ & $0.455464-0.059655 i$ & $0.455464-0.059655 i$ & $0\%$\\
$0.12$ & $0.102151-0.056019 i$ & $0.103227-0.055849 i$ & $0.935\%$ & $0.24$ & $0.291741-0.052606 i$ & $0.291741-0.052606 i$ & $<10^{-4}\%$ & $0.24$ & $0.458039-0.059034 i$ & $0.458039-0.059034 i$ & $0\%$\\
$0.14$ & $0.102719-0.052109 i$ & $0.101917-0.052457 i$ & $0.759\%$ & $0.26$ & $0.296058-0.050440 i$ & $0.296058-0.050440 i$ & $0.00003\%$ & $0.26$ & $0.460853-0.058345 i$ & $0.460853-0.058345 i$ & $0\%$\\
$0.16$ & $0.105552-0.047077 i$ & $0.105141-0.046968 i$ & $0.367\%$ & $0.28$ & $0.300717-0.047973 i$ & $0.300717-0.047973 i$ & $0.0002\%$ & $0.28$ & $0.463909-0.057586 i$ & $0.463909-0.057586 i$ & $0\%$\\
 &  &  &  & $0.3$ & $0.305705-0.045171 i$ & $0.305703-0.045172 i$ & $0.0005\%$ & $0.3$ & $0.467212-0.056750 i$ & $0.467212-0.056750 i$ & $0\%$\\
 &  &  &  & $0.32$ & $0.310992-0.042003 i$ & $0.310993-0.042005 i$ & $0.00075\%$ & $0.32$ & $0.470765-0.055833 i$ & $0.470765-0.055833 i$ & $0\%$\\
 &  &  &  & $0.34$ & $0.316555-0.038447 i$ & $0.316566-0.038463 i$ & $0.0061\%$ & $0.34$ & $0.474574-0.054830 i$ & $0.474574-0.054830 i$ & $0\%$\\
 &  &  &  & $0.36$ & $0.322345-0.034530 i$ & $0.322349-0.034577 i$ & $0.0147\%$ & $0.36$ & $0.478644-0.053733 i$ & $0.478644-0.053733 i$ & $0\%$\\
 &  &  &  & $0.38$ & $0.328200-0.030527 i$ & $0.328377-0.030882 i$ & $0.120\%$ & $0.38$ & $0.482981-0.052534 i$ & $0.482981-0.052534 i$ & $0\%$\\
 &  &  &  &  &  &  &  & $0.4$ & $0.487591-0.051224 i$ & $0.487591-0.051224 i$ & $0\%$\\
 &  &  &  &  &  &  &  & $0.42$ & $0.492479-0.049793 i$ & $0.492479-0.049793 i$ & $0\%$\\
 &  &  &  &  &  &  &  & $0.44$ & $0.497653-0.048228 i$ & $0.497653-0.048228 i$ & $0\%$\\
 &  &  &  &  &  &  &  & $0.46$ & $0.503117-0.046514 i$ & $0.503117-0.046514 i$ & $0\%$\\
 &  &  &  &  &  &  &  & $0.48$ & $0.508878-0.044635 i$ & $0.508878-0.044635 i$ & $0\%$\\
 &  &  &  &  &  &  &  & $0.5$ & $0.514939-0.042571 i$ & $0.514939-0.042571 i$ & $0\%$\\
 &  &  &  &  &  &  &  & $0.52$ & $0.521302-0.040299 i$ & $0.521302-0.040299 i$ & $0\%$\\
 &  &  &  &  &  &  &  & $0.54$ & $0.527966-0.037794 i$ & $0.527966-0.037794 i$ & $0\%$\\
 &  &  &  &  &  &  &  & $0.56$ & $0.534924-0.035031 i$ & $0.534924-0.035033 i$ & $0.00041\%$\\
 &  &  &  &  &  &  &  & $0.58$ & $0.542166-0.031985 i$ & $0.542164-0.031988 i$ & $0.00066\%$\\
 &  &  &  &  &  &  &  & $0.6$ & $0.549653-0.028680 i$ & $0.549652-0.028683 i$ & $0.00049\%$\\
 &  &  &  &  &  &  &  & $0.62$ & $0.557691-0.025642 i$ & $0.558147-0.025823 i$ & $0.0878\%$\\
\hline
\end{tabular}%
}
\caption{Massive scalar quasinormal modes for fixed $l_0=0.8$ ($M=1$).  The table combines the $\ell=0,1,2$ data obtained from the 16th- and 14th-order Pad\'e-resummed WKB approximants.  The last column in each block gives the relative difference in percent.}
\label{tab:qnm-l008}

\end{table*}

\begin{figure*}
\centering
\includegraphics[width=0.98\textwidth]{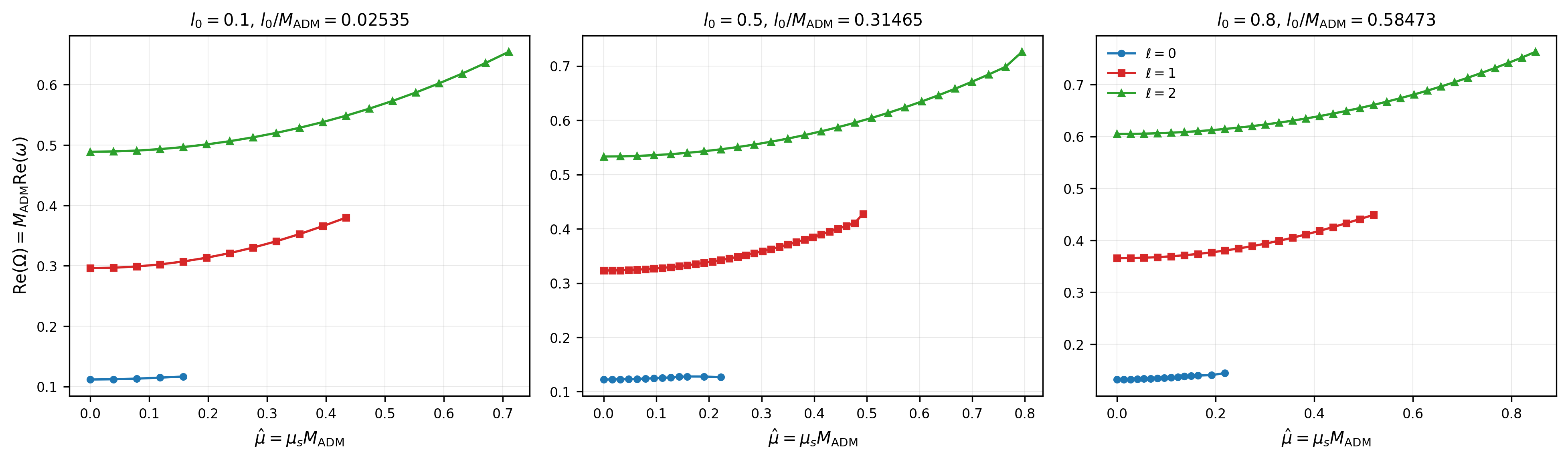}
\caption{Real parts of the massive scalar quasinormal frequencies in ADM units.  The horizontal axis is the ADM-scaled scalar mass $\hat\mu=\mu_sM_{\rm ADM}$ and the vertical axis is ${\rm Re}(\Omega)=M_{\rm ADM}{\rm Re}(\omega)$.  The three panels correspond to $l_0=0.1$, $0.5$, and $0.8$ with $M=1$, or equivalently $l_0/M_{\rm ADM}=0.02535$, $0.31465$, and $0.58473$.  The plotted points are the WKB16 entries satisfying the reliability criterion $\Delta\leq1\%$.}
\label{fig:massive-scalar-qnm-adm-real}
\end{figure*}

\begin{figure*}
\centering
\includegraphics[width=0.98\textwidth]{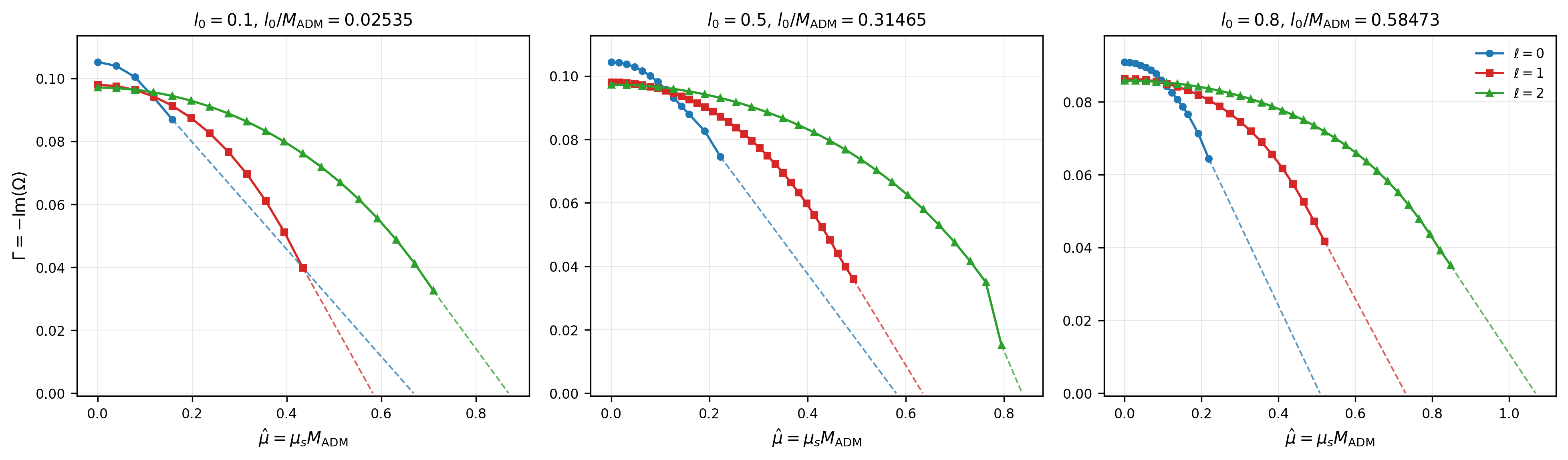}
\caption{Damping rates of the massive scalar quasinormal modes in ADM units.  The plotted quantity is $\Gamma=-{\rm Im}(\Omega)$ as a function of $\hat\mu=\mu_sM_{\rm ADM}$.  Solid curves show the reliable WKB16 data, while dashed segments give a linear extrapolation of the last three reliable points to $\Gamma=0$.  The extrapolation indicates the quasiresonant tendency of the massive scalar modes and is not used as an additional WKB data point.}
\label{fig:massive-scalar-qnm-adm-damping}
\end{figure*}

The entries in Tables~\ref{tab:qnm-l001}--\ref{tab:qnm-l008} are converted to ADM variables through
\begin{equation}\label{adm-rescaling-qnm}
\begin{aligned}
 \hat\mu&=\mu_sM_{\rm ADM},
 &\Omega&=M_{\rm ADM}\omega,\\
 M_{\rm ADM}&=1+\frac{3\pi}{32l_0}
 & &\quad (M=1).
\end{aligned}
\end{equation}
Figures~\ref{fig:massive-scalar-qnm-adm-real} and \ref{fig:massive-scalar-qnm-adm-damping} show the same WKB16 data in these physical variables.  The conversion is important because changing $l_0$ also changes the ADM mass when the bare parameter $M$ is kept fixed.

\begin{figure*}
\centering
\includegraphics[width=0.98\textwidth]{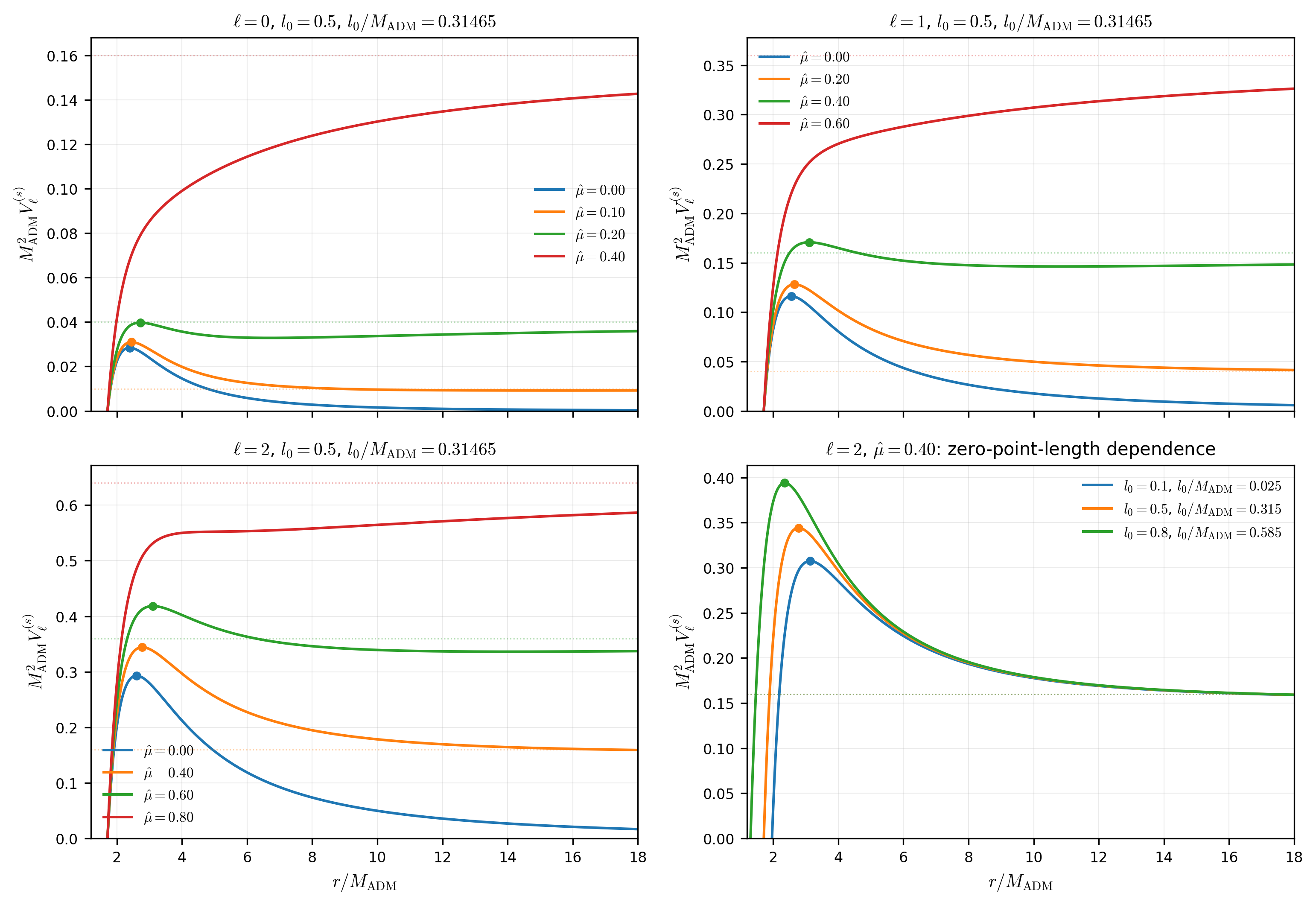}
\caption{Typical ADM-scaled massive-scalar effective potentials.  The first three panels use $l_0=0.5$ and show how the local barrier evolves with $\hat\mu=\mu_sM_{\rm ADM}$ for $\ell=0,1,2$.  Dotted horizontal lines mark the asymptotic plateaus $M_{\rm ADM}^2V_\ell^{(s)}\to\hat\mu^2$, and filled circles mark local maxima when they are present.  The last curve in each of these panels illustrates the loss of a clean potential peak.  The lower-right panel fixes $\ell=2$ and $\hat\mu=0.40$ and compares different zero-point lengths; increasing $l_0/M_{\rm ADM}$ moves the barrier inward and raises its peak in the displayed range.}
\label{fig:massive-scalar-effective-potentials}
\end{figure*}

Several trends are visible directly from the ADM-scaled figures.  First, ${\rm Re}(\Omega)$ increases with $\hat\mu$ for all displayed branches.  This is the expected behavior for a massive field because the mass term raises the asymptotic value of the potential and therefore increases the oscillation scale.  Second, the damping rate decreases as the scalar mass grows.  The decrease is monotonic on the reliable branches shown in Fig.~\ref{fig:massive-scalar-qnm-adm-damping}; it is fastest near the end of the branch, where the massive field approaches a long-lived, threshold-like oscillation.  Third, the allowed WKB interval depends strongly on the multipole.  The $\ell=0$ branch terminates earliest, while the centrifugal barrier allows the $\ell=2$ modes to remain stable over a wider interval of $\hat\mu$.

The potential profiles in Fig.~\ref{fig:massive-scalar-effective-potentials} explain this termination of the WKB branches.  For small $\hat\mu$ the curvature and centrifugal terms produce a local maximum outside the horizon, so the WKB expansion can be organized around a barrier peak.  Increasing $\hat\mu$ raises the asymptotic plateau and changes the outer part of the potential; after a multipole-dependent value of the scalar mass the maximum and the outer dip are washed out, leaving no suitable single peak for the barrier WKB calculation.  This happens first for $\ell=0$ and later for higher multipoles because the centrifugal term $\ell(\ell+1)/r^2$ keeps the barrier more pronounced.

The relation between the barrier height and the real part of the frequency should therefore be stated with some care.  The leading term in Eq.~(\ref{WKBformula}) ties ${\rm Re}(\omega)^2$ to the peak value $V_0$, so along a fixed, well-defined barrier branch a taller peak usually corresponds to a larger oscillation frequency.  The numerical data confirm this trend for the $\ell=1,2$ branches and for most of the monopole points.  In the massive case, however, $\hat\mu$ also changes the asymptotic plateau and the higher derivatives entering the WKB corrections; consequently the peak height alone is not a complete predictor.  This is why the statement is safest in the local sense: while the barrier remains clean, raising the potential scale generally raises ${\rm Re}(\omega)$ and lowers the damping, but close to peak disappearance the WKB branch becomes delicate and small nonmonotonicities can appear.

To quantify the tendency toward zero damping, we fit the last three reliable points of each branch by a straight line in the $(\hat\mu,\Gamma)$ plane and determine where that line reaches $\Gamma=0$.  The resulting values, denoted by $\hat\mu_c^{\rm lin}$, are listed in Table~\ref{tab:critical-mu}.  They should be interpreted as local indicators of the quasiresonant trend rather than as exact eigenfrequencies, because the WKB barrier approximation becomes increasingly delicate close to the point where the potential peak disappears.

\begin{table}
\centering
\scriptsize
\setlength{\tabcolsep}{4pt}
\begin{tabular}{c c c c c}
\hline
$l_0$ & $l_0/M_{\rm ADM}$ & $\ell$ & $\hat\mu_{\rm last}$ & $\hat\mu_c^{\rm lin}$\\
\hline
$0.1$ & $0.02535$ & $0$ & $0.15781$ & $0.66826$\\
$0.1$ & $0.02535$ & $1$ & $0.43398$ & $0.58208$\\
$0.1$ & $0.02535$ & $2$ & $0.71014$ & $0.86919$\\
$0.5$ & $0.31465$ & $0$ & $0.22247$ & $0.58025$\\
$0.5$ & $0.31465$ & $1$ & $0.49261$ & $0.63444$\\
$0.5$ & $0.31465$ & $2$ & $0.79452$ & $0.83662$\\
$0.8$ & $0.58473$ & $0$ & $0.21890$ & $0.50817$\\
$0.8$ & $0.58473$ & $1$ & $0.51990$ & $0.73094$\\
$0.8$ & $0.58473$ & $2$ & $0.84826$ & $1.06911$\\
\hline
\end{tabular}
\caption{Indicative ADM-scaled scalar masses at which the damping rate tends to zero.  Here $\hat\mu_{\rm last}$ is the largest reliable scalar mass shown in the corresponding WKB branch, and $\hat\mu_c^{\rm lin}$ is estimated by linearly extrapolating the last three reliable values of $\Gamma=-{\rm Im}(\Omega)$ to zero.}
\label{tab:critical-mu}
\end{table}

The extrapolated values show that the zero-damping tendency occurs at finite ADM-scaled scalar mass.  For the higher multipoles the last reliable WKB points already lie close to the extrapolated critical region; for example, at $l_0=0.5$ and $\ell=2$ the last point is at $\hat\mu=0.79452$ and the linear zero is at $\hat\mu_c^{\rm lin}=0.83662$.  The $\ell=0$ estimates are less precise, because this branch loses a clean barrier earlier and the extrapolation covers a wider interval.  Nevertheless, the same qualitative behavior is present: increasing the scalar mass suppresses the imaginary part and drives the mode toward a long-lived configuration.

The dependence on the zero-point length is also visible.  At fixed $M=1$, increasing $l_0$ decreases $M_{\rm ADM}$ and moves the curves to different ranges of $\hat\mu$.  In ADM units the $l_0=0.8$ case corresponds to the largest deformation considered here, $l_0/M_{\rm ADM}=0.58473$, and it has smaller massless damping rates than the weakly deformed case.  The self-energy deformation therefore changes both the starting point of the branch and the scalar mass at which the quasiresonant trend becomes visible.

For completeness, Table~\ref{tab:qnm-l005-ell2-overtones} displays the first two overtones for the same representative $l_0=0.5$, $\ell=2$ sector.  We keep only the entries for which a potential peak is present and the relative difference between the 16th and 14th-order Pad\'e approximants does not exceed one percent.

\begin{widetext}
\begin{center}
\refstepcounter{table}\label{tab:qnm-l005-ell2-overtones}
\scriptsize
\setlength{\tabcolsep}{2.0pt}
\renewcommand{\arraystretch}{0.9}
\resizebox{\textwidth}{!}{%
\begin{tabular}{c c c c c c c}
\hline
 & \multicolumn{3}{c}{$n=1$} & \multicolumn{3}{c}{$n=2$}\\
\hline
$\mu_s$ & WKB16 & WKB14 & diff. & WKB16 & WKB14 & diff.\\
\hline
$0$ & $0.325205-0.186379 i$ & $0.325205-0.186379 i$ & $0\%$ & $0.307639-0.318607 i$ & $0.307639-0.318607 i$ & $0\%$\\
$0.02$ & $0.325302-0.186260 i$ & $0.325302-0.186260 i$ & $0\%$ & $0.307659-0.318492 i$ & $0.307659-0.318492 i$ & $0\%$\\
$0.04$ & $0.325592-0.185902 i$ & $0.325592-0.185902 i$ & $0\%$ & $0.307720-0.318146 i$ & $0.307720-0.318146 i$ & $0\%$\\
$0.06$ & $0.326075-0.185303 i$ & $0.326075-0.185303 i$ & $0\%$ & $0.307819-0.317569 i$ & $0.307819-0.317569 i$ & $0\%$\\
$0.08$ & $0.326750-0.184464 i$ & $0.326750-0.184464 i$ & $0\%$ & $0.307955-0.316762 i$ & $0.307955-0.316762 i$ & $0\%$\\
$0.1$ & $0.327616-0.183380 i$ & $0.327616-0.183380 i$ & $0\%$ & $0.308125-0.315725 i$ & $0.308125-0.315725 i$ & $<10^{-4}\%$\\
$0.12$ & $0.328671-0.182050 i$ & $0.328671-0.182050 i$ & $0\%$ & $0.308327-0.314460 i$ & $0.308327-0.314460 i$ & $<10^{-4}\%$\\
$0.14$ & $0.329911-0.180470 i$ & $0.329911-0.180470 i$ & $0\%$ & $0.308557-0.312968 i$ & $0.308556-0.312969 i$ & $0.0002\%$\\
$0.16$ & $0.331335-0.178635 i$ & $0.331335-0.178635 i$ & $0\%$ & $0.308810-0.311251 i$ & $0.308809-0.311251 i$ & $0.0002\%$\\
$0.18$ & $0.332938-0.176542 i$ & $0.332938-0.176542 i$ & $0\%$ & $0.309080-0.309311 i$ & $0.309080-0.309311 i$ & $0.0002\%$\\
$0.2$ & $0.334716-0.174184 i$ & $0.334716-0.174184 i$ & $0\%$ & $0.309363-0.307151 i$ & $0.309363-0.307151 i$ & $0\%$\\
$0.22$ & $0.336661-0.171556 i$ & $0.336661-0.171556 i$ & $0\%$ & $0.309652-0.304775 i$ & $0.309652-0.304774 i$ & $0.0002\%$\\
$0.24$ & $0.338768-0.168652 i$ & $0.338768-0.168652 i$ & $0\%$ & $0.309941-0.302188 i$ & $0.309942-0.302186 i$ & $0.0005\%$\\
$0.26$ & $0.341027-0.165466 i$ & $0.341027-0.165466 i$ & $0\%$ & $0.310222-0.299394 i$ & $0.310223-0.299391 i$ & $0.0008\%$\\
$0.28$ & $0.343426-0.161990 i$ & $0.343426-0.161990 i$ & $0\%$ & $0.310497-0.296402 i$ & $0.310498-0.296402 i$ & $0.0002\%$\\
$0.3$ & $0.345954-0.158221 i$ & $0.345954-0.158221 i$ & $0\%$ & $0.310748-0.293223 i$ & $0.310758-0.293237 i$ & $0.0040\%$\\
$0.32$ & $0.348595-0.154152 i$ & $0.348595-0.154152 i$ & $0.0002\%$ & $0.310975-0.289862 i$ & $0.310979-0.289892 i$ & $0.0072\%$\\
$0.34$ & $0.351332-0.149782 i$ & $0.351329-0.149782 i$ & $0.0008\%$ & $0.311179-0.286345 i$ & $0.311177-0.286365 i$ & $0.0047\%$\\
$0.36$ & $0.354144-0.145111 i$ & $0.354140-0.145109 i$ & $0.0011\%$ & $0.311335-0.282687 i$ & $0.311334-0.282689 i$ & $0.0007\%$\\
$0.38$ & $0.357009-0.140139 i$ & $0.357009-0.140139 i$ & $0.0001\%$ & $0.311467-0.278859 i$ & $0.311390-0.278944 i$ & $0.0275\%$\\
$0.4$ & $0.359912-0.134878 i$ & $0.359905-0.134844 i$ & $0.0090\%$ & $0.311510-0.274939 i$ & $0.311104-0.275079 i$ & $0.103\%$\\
$0.42$ & $0.362847-0.129287 i$ & $0.362972-0.129249 i$ & $0.0341\%$ & $0.311079-0.270950 i$ & $0.311070-0.270945 i$ & $0.0024\%$\\
\hline
\end{tabular}%
}
\par\vspace{3pt}
\begin{minipage}{0.95\textwidth}
\small TABLE~\thetable. Massive scalar quasinormal overtones for fixed $l_0=0.5$, $\ell=2$, and $M=1$.  The table combines the $n=1$ and $n=2$ data obtained from the 16th- and 14th-order Pad\'e-resummed WKB approximants.  Only entries in the common reliable interval are retained: a potential peak must be present and the relative difference must satisfy $\Delta\leq1\%$ for both displayed overtones.
\end{minipage}
\end{center}
\end{widetext}

\begin{figure*}
\refstepcounter{figure}
\includegraphics[width=0.98\textwidth]{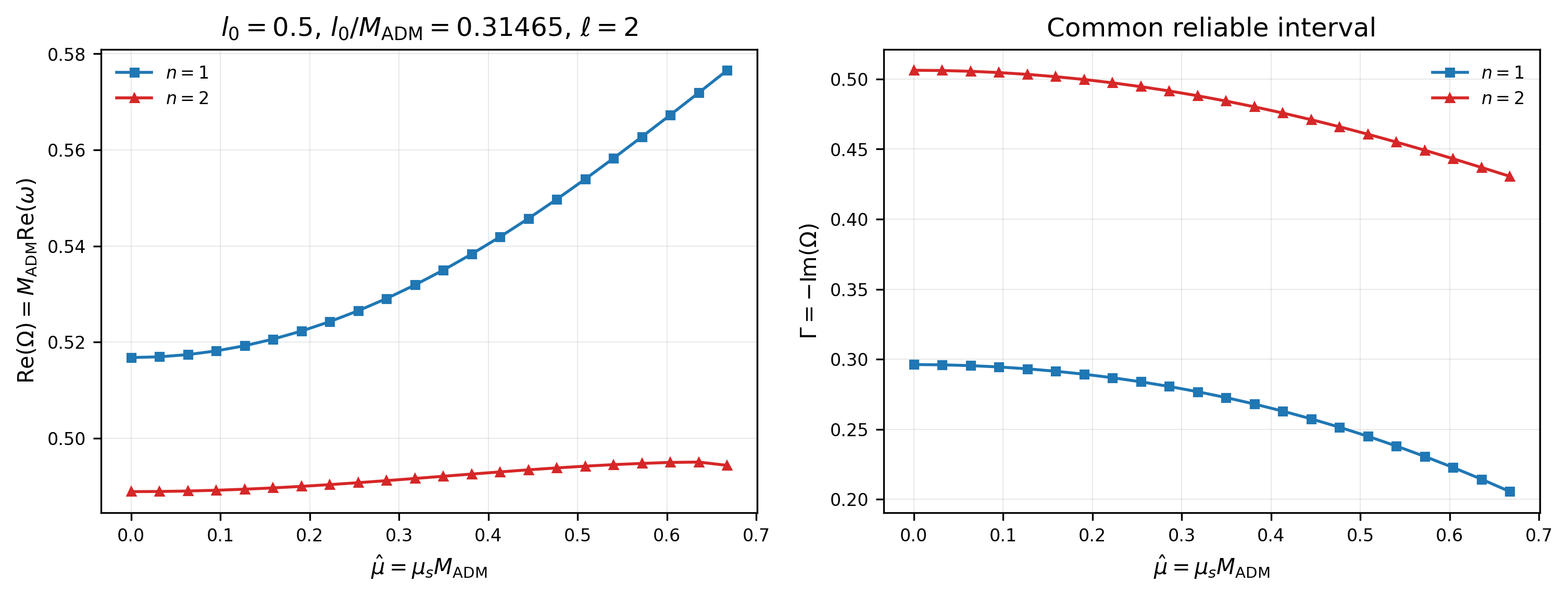}
\small FIG.~\thefigure. ADM-scaled first and second overtones for the $l_0=0.5$, $\ell=2$ scalar branch.  The plotted points are the WKB16 entries from Table~\ref{tab:qnm-l005-ell2-overtones} after applying the same common-interval filtering used in the table.
\end{figure*}

The overtone spectra in the accompanying plot follow the same qualitative mass dependence as the fundamental branch, but with the expected stronger damping.  At fixed $l_0=0.5$ and $\ell=2$, the $n=2$ mode has a smaller real part and a larger damping rate than the $n=1$ mode over the displayed common reliable interval.  Increasing $\hat\mu$ raises the oscillation frequency and decreases $\Gamma=-{\rm Im}(\Omega)$ for both overtones, showing that the approach toward long-lived massive oscillations is not restricted to the fundamental mode.

The higher overtone is also more sensitive to the loss of WKB reliability.  To compare the two overtones on equal footing, Table~\ref{tab:qnm-l005-ell2-overtones} and the accompanying plot stop at the last mass for which both branches remain inside the same reliability window.  Beyond this endpoint the massive plateau and the barrier maximum cease to provide a clean single-peak WKB problem for the overtone comparison.

\section{Geodesic observables in ADM units}\label{sec:geodesic-results}

We now evaluate the photon-ring, ISCO and shadow observables for ten values of $l_0$ uniformly spanning $0.1\leq l_0\leq0.85$.  The results are given in Table~\ref{tab:geodesic-observables}.  As in the QNM section, the bare mass is set to $M=1$ during the calculation.  The table lists both the sampled bare value of $l_0$ and the corresponding deformation $l_0/M_{\rm ADM}$; all radii, frequencies and Lyapunov exponents are then reported in ADM units.

\begin{widetext}
\begin{center}
\refstepcounter{table}\label{tab:geodesic-observables}
\scriptsize
\setlength{\tabcolsep}{4pt}
\resizebox{\textwidth}{!}{%
\begin{tabular}{c c c c c c c c c}
\hline
$l_0$ & $l_0/M_{\rm ADM}$ & $r_+/M_{\rm ADM}$ & $r_{\rm ph}/M_{\rm ADM}$ & $M_{\rm ADM}\Omega_{\rm ph}$ & $M_{\rm ADM}\lambda$ & $R_{\rm sh}/M_{\rm ADM}$ & $r_{\rm ISCO}/M_{\rm ADM}$ & $E_b$\\
\hline
$0.100$ & $0.02535$ & $1.96722$ & $2.95640$ & $0.19457$ & $0.19312$ & $5.13966$ & $5.90225$ & $0.05806$\\
$0.183$ & $0.07034$ & $1.92198$ & $2.89677$ & $0.19751$ & $0.19390$ & $5.06307$ & $5.76999$ & $0.05927$\\
$0.267$ & $0.12671$ & $1.87355$ & $2.83403$ & $0.20065$ & $0.19446$ & $4.98388$ & $5.63334$ & $0.06057$\\
$0.350$ & $0.19006$ & $1.82245$ & $2.76952$ & $0.20389$ & $0.19462$ & $4.90451$ & $5.49630$ & $0.06192$\\
$0.433$ & $0.25799$ & $1.76641$ & $2.70099$ & $0.20735$ & $0.19423$ & $4.82287$ & $5.35514$ & $0.06336$\\
$0.517$ & $0.32908$ & $1.70193$ & $2.62496$ & $0.21117$ & $0.19309$ & $4.73547$ & $5.20387$ & $0.06497$\\
$0.600$ & $0.40245$ & $1.62412$ & $2.53697$ & $0.21561$ & $0.19085$ & $4.63802$ & $5.03539$ & $0.06683$\\
$0.683$ & $0.47752$ & $1.52487$ & $2.43075$ & $0.22099$ & $0.18685$ & $4.52507$ & $4.84097$ & $0.06909$\\
$0.767$ & $0.55389$ & $1.38586$ & $2.29571$ & $0.22787$ & $0.17961$ & $4.38837$ & $4.60840$ & $0.07195$\\
$0.850$ & $0.63127$ & $1.10449$ & $2.10737$ & $0.23742$ & $0.16459$ & $4.21201$ & $4.31694$ & $0.07579$\\
\hline
\end{tabular}%
}
\par\vspace{3pt}
\begin{minipage}{0.95\textwidth}
\small TABLE~\thetable. Particle-motion and optical observables in ADM units.  The ten rows use uniformly spaced values of $l_0$ between $0.1$ and $0.85$.  Here $r_+$ is the event-horizon radius, $r_{\rm ph}$ is the unstable null circular orbit, $\Omega_{\rm ph}$ is its orbital frequency, $\lambda$ is its Lyapunov exponent, $R_{\rm sh}$ is the shadow radius, $r_{\rm ISCO}$ is the innermost stable circular orbit and $E_b=1-E(r_{\rm ISCO})$ is the ISCO binding energy.
\end{minipage}
\end{center}
\end{widetext}
Several conclusions follow from these numbers.  First, both characteristic radii decrease when $l_0/M_{\rm ADM}$ grows: $r_{\rm ph}/M_{\rm ADM}$ changes from $2.95640$ to $2.10737$, while $r_{\rm ISCO}/M_{\rm ADM}$ changes from $5.90225$ to $4.31694$.  Thus the self-energy deformation makes the exterior potential structure more compact in physical units.  Second, the photon-ring frequency increases from $M_{\rm ADM}\Omega_{\rm ph}=0.19457$ to $0.23742$.  This trend is consistent with the inward motion of the photon ring and with the eikonal expectation that a more compact photon ring oscillates faster.  Third, the shadow radius decreases from $R_{\rm sh}/M_{\rm ADM}=5.13966$ to $4.21201$, giving a direct optical counterpart of the same effect.

The Lyapunov exponent is less monotonic.  It grows slightly at small deformation, reaches $M_{\rm ADM}\lambda=0.19462$ near $l_0=0.350$, and then decreases to $0.16459$ at $l_0=0.850$.  This means that the photon-ring instability time is not controlled only by the orbital radius; it also depends on the second derivative of the optical potential at the null orbit.  The ISCO binding energy, by contrast, grows monotonically from $E_b=0.05806$ to $0.07579$.  The regular self-energy correction therefore increases the nominal accretion efficiency in this parameter range.

A possible observational interpretation is to compare the dimensionless shadow radius with the EHT angular-diameter measurements.  This comparison must be treated only as an order-of-magnitude constraint, because EHT images measure an emission ring whose relation to the vacuum critical curve depends on the mass-to-distance ratio, spin, inclination and plasma model.  Nevertheless, the measured M87* ring diameter, $42\pm3\,\mu{\rm as}$, and the Sgr~A* metric tests are consistent with a Kerr shadow at the ten-percent level~\cite{EventHorizonTelescope:2019dse,EventHorizonTelescope:2022xqj}.  In the present static model the Schwarzschild value is $R_{\rm sh}/M_{\rm ADM}=3\sqrt{3}=5.196$, while Table~\ref{tab:geodesic-observables} gives $R_{\rm sh}/M_{\rm ADM}=5.13966$--$4.21201$.  Equivalently, the fractional deviation $\delta_{\rm sh}=R_{\rm sh}/(3\sqrt{3}M_{\rm ADM})-1$ runs from $-0.0109$ to $-0.189$. Thus a purely illustrative $|\delta_{\rm sh}|\lesssim0.1$ shadow-size requirement would keep roughly $l_0\lesssim0.55$ in the present $M=1$ sampling, or $l_0/M_{\rm ADM}\lesssim0.35$, whereas the largest deformation in the table would be mildly disfavored.

These particle observables provide a useful counterpart to the scalar spectra.  The massive scalar modes are sensitive to the wave potential $V_\ell^{(s)}$, including the asymptotic mass plateau, whereas the geodesic quantities are determined solely by the metric function and its first two derivatives at special circular orbits.  Both sets of quantities therefore respond to the same deformation of $f(r)$, but they emphasize different regions of the effective-potential landscape: the WKB modes follow the barrier peak of the field equation, the shadow follows the maximum of $f/r^2$, and the ISCO follows the marginal point of the timelike potential.

\section{Conclusions}\label{sec:conclusions}

We have analyzed wave and particle probes of the regular T-duality-inspired black hole with gravitational self-energy.  On the wave side, the calculation uses the massive Klein--Gordon potential in Eq.~(\ref{scalar-potential}) and high-order Pad\'e-resummed WKB approximants.  On the particle side, it uses the null and timelike geodesic potentials in Eq.~(\ref{geodesic-potential}).  The physical parameter space is organized by $l_0/M_{\rm ADM}$ for the geometry and by $\hat\mu=\mu_sM_{\rm ADM}$ for the scalar-field mass.

The massive scalar data show a consistent pattern.  Along every reliable WKB branch, the ADM-scaled oscillation frequency ${\rm Re}(\Omega)$ increases as $\hat\mu$ grows, whereas the damping rate $\Gamma=-{\rm Im}(\Omega)$ decreases.  This decrease is the clearest signature of the approach to quasiresonant behavior.  Linear extrapolations of the last reliable points indicate finite values of $\hat\mu$ at which $\Gamma$ tends to zero.  The precise critical value depends on the multipole and on $l_0/M_{\rm ADM}$, and the $\ell=0$ estimates are less robust because the potential barrier disappears earlier for the lowest multipole.

The particle-motion observables give an independent view of the same deformation.  Over the interval $0.1\leq l_0\leq0.85$ with $M=1$, the ADM-scaled photon-ring radius decreases from $2.95640$ to $2.10737$, the shadow radius decreases from $5.13966$ to $4.21201$, and the photon-ring frequency increases from $0.19457$ to $0.23742$.  The ISCO also moves inward, from $r_{\rm ISCO}/M_{\rm ADM}=5.90225$ to $4.31694$, while the binding energy grows from $0.05806$ to $0.07579$.  These trends show that the regular self-energy correction makes the strong-field region more compact in ADM units and increases the nominal accretion efficiency.

The common concept behind the two analyses is the deformation of effective potentials.  The massive scalar modes follow the peak of the field-theory barrier and its evolution as the asymptotic mass plateau is raised.  The photon-ring and shadow observables follow the maximum of $f(r)/r^2$, while the ISCO follows the marginal-stability point of the timelike potential.  Expressing all quantities in ADM units separates these effects from the trivial mass rescaling and makes the comparison between wave and particle probes transparent.

The quasinormal frequencies obtained here may also be used as input for approximate estimates of grey-body factors.  Recent work has shown that, for black-hole effective potentials with the appropriate barrier structure, there is a correspondence between grey-body factors and quasinormal-mode data~\cite{Konoplya:2024vuj,Konoplya:2024lir,Bolokhov:2024otn}.  Applying this relation to the present massive-scalar spectra would provide an economical way to infer transmission probabilities for the regular self-energy geometry, complementing a direct scattering calculation.

A more detailed treatment near the critical scalar masses should use direct integration or spectral methods for the massive-field quasinormal boundary-value problem, because the barrier WKB approximation becomes less reliable as the peak disappears.  On the particle side, the next step would be to include finite-size disk observables or ray-traced images, which would connect the shadow radius and ISCO quantities computed here to synthetic electromagnetic signatures of the same geometry.

\section*{Declaration of Competing Interest}
The author declares that he has no known competing financial interests or personal relationships that could have appeared to influence the work reported in this paper.

\section*{Data Availability}
The numerical data used to generate the figures are displayed in the tables and are provided in the accompanying data files and plotting scripts in this manuscript folder.

\begin{acknowledgments}
The author acknowledges the University of Seville for their support through the Plan-US of aid to Ukraine.
\end{acknowledgments}

\bibliography{MassiveScalarSelfEnergyBH}

\end{document}